\documentclass[journal]{IEEEtran}
\usepackage{cite}
\usepackage{amsmath,amssymb,amsfonts}
\usepackage{algorithmic}
\usepackage{graphicx}
\usepackage{textcomp}
\usepackage{hyperref}

\usepackage{multirow}
\usepackage{booktabs,siunitx}
\usepackage{makecell}

\usepackage{balance}
\usepackage{flushend}

\begin{document}
\title{Design of Adaptive Backstepping Control for Direct Power Control of Three-Phase PWM Rectifier)}
\author{
	Basit~Olakunle Alawode\textsuperscript{1}, Sami El Ferik\textsuperscript{2}\\
	Electrical Engineering Department, \\
	King Fahd University of Petroleum and Minerals, Dhahran, Saudi Arabia. \\
	Email: {g201707310@kfupm.edu.sa\textsuperscript{1}, selferik@kfupm.edu.sa\textsuperscript{2}}
}
\maketitle

\begin{abstract} 
In this paper, we focused on the design of an adaptive backstepping controller (adaptive-BSC) for direct power control (DPC) of a three-phase PWM rectifier. In the proposed system, it is desired to control both the output DC voltage of the rectifier and the reactive power simultaneously by making them track desired respective values. This was done by having independent virtual control signals for both the output voltage and the reactive power. The adaptive control signals were gotten from the dynamic equations of the three-phase system. For comparison, both the BSC and adaptive-BSC equations were developed. Numerical simulations were performed on both of them on a 5kW system. The proposed adaptive-BSC was designed to work under more challenging system variations as compared with the BSC as it has to estimate the value of the unknown system load. Despite this, it still performed better than its BSC counterpart.  	
\end{abstract}

\begin{IEEEkeywords}  
	Three-phase Rectifiers, PWM Rectifier, Direct Power Control, Adaptive Control, Adaptive Backstepping Control, Backstepping Control, Instantaneous Power, Reactive Power.
\end{IEEEkeywords}

\section{Introduction}
\textit{Rectification} is simply the process by which alternating current (AC) is being converted to direct current (DC) with the help of rectifiers. For an uncontrolled rectification, where the entirety of the input power is desired at the output, power diodes are used for the rectification. For controlled rectification, power switches are used instead. The operation of the switches are controlled using pulse width modulation (PWM) generated signals. The PWM controlled rectifiers, which are operated in four-quadrants \cite{kant}, help in controlling how much DC power reaches the output from the AC input. Three-phase PWM controlled rectifiers are widely used in industries, micro-grids, renewable energy generation system like wind power generation. With increase in using renewable energy sources for power generation and micro-grid implementation, storage units are required. Electric vehicles (EVs) can be used as storage which can help feed energy back into the grid through vehicle-to-grid (V2G) technology \cite{main2, main3}. To store energy or charge up EVs, three-phase rectifiers are also needed. 

For typical applications highlighted above, control of output voltage and power factor is essentially not enough. Control must be able to respond swiftly in order to meet the demands of such applications. There are majorly two types of control approaches being employed for three-phase PWM rectifiers: direct power control (DPC) and voltage-oriented control (VOC). The VOC is based on the current controller which are very sensitive to variations and external disturbances \cite{main4}. But, to be able to use rectifiers for the above purposes, it must be immune to external variations and disturbances. The DPC approach helps in solving this issue. It is able to do so because it is based on using the instantaneous active and reactive power theory which has a better performance on the dynamic response \cite{main5, main6}. Because of its ability to give better performance on dynamic response without current control loop, the DPC has been used extensively in the control of three-phase PWM rectifier systems \cite{dpc1, dpc2, dpc3, dpc4}.   

The backstepping control for DPC of three-phase PWM rectifier has been proposed by Rong, et al. \cite{main} in 2018. This was used to control the reactive and output voltage of the rectifier. This system, however, breaks down in varying load conditions. Backstepping control has also found applications in non-linear feedback systems \cite{main8, main7}. In this paper, we take a step further by proposing an adaptive backstepping control (adaptive-BSC) for the same system proposed in \cite{main}. This ensures that the system is robust and able to adapt to variations and external disturbances, especially in the presence of varying load. Due to its ability to adapt to variations and external disturbances, adaptive backstepping has also been used in vast application areas containing huge system uncertainties \cite{adap1,adap2,adap3,adap4,adap5,adap6}.

The remainder of this paper is presented as follows: Section II gives the model of the three-phase rectifier. Section III discusses the backstepping control of the reactive power and output voltage. Section IV focusses on the adaptive backstepping controller design while section V discusses the results and simulations performed. The paper concludes in section VI.
       
\section{Three-Phase PWM Rectifier Model}
\begin{figure}[htbp!]
	\centering
	\includegraphics[width=0.45\textwidth]{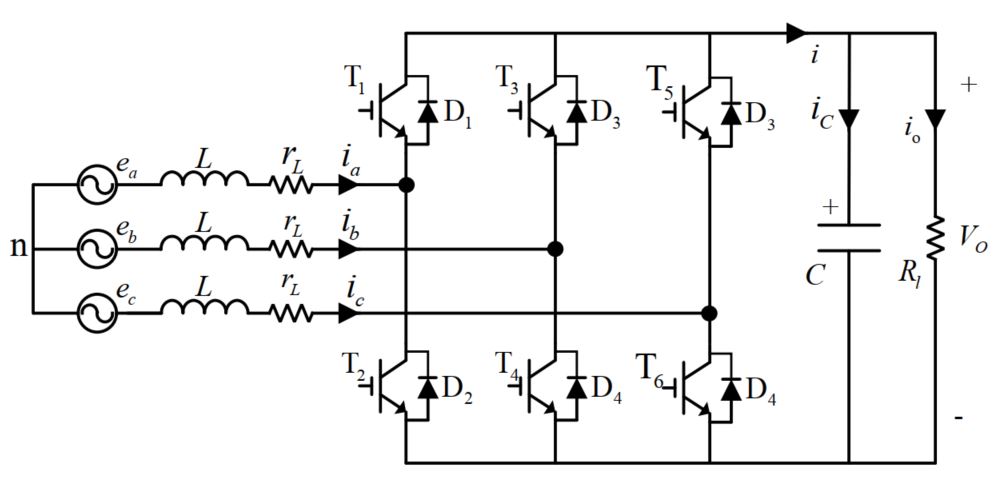}
	\caption[]{Three Phase PWM Rectifier Circuit \cite{main}}
	\label{fig:ckt}
\end{figure}

A three-phase PWM rectifier circuit is as shown in Fig. \ref{fig:ckt}. This is a Y connected voltage source system where $e_{a}$, $e_{b}$ and $e_{c}$ are the voltages of the \textit{red, yellow} and \textit{blue} phases respectively. $n$ is the neutral point. The source voltages are assumed to be balanced which means that they are $120^{\circ}$ out of phase with each other. The source is filtered with the help of inductor \textit{L} whose equivalent resistance is represented by $r_{L}$. The current flowing through each phase is denoted by $i_{a}$, $i_{b}$ and $i_{c}$ respectively. Power switches $T_{1}$ to $T_{6}$ are used to achieve the rectification with the help of a controlled PWM signals to their gate terminals. Capacitor \textit{C} is used to filter the output DC voltage $V_{o}$ and the load is a varying $R_{l}$ resistive load.

For ease of analysis, the synchronous \textit{abc} frame source voltages ($e_{a}$, $e_{b}$, $e_{c}$) and currents ($i_{a}$, $i_{b}$, $i_{c}$), are projected to the stationary \text{d-q} frame resulting in $e_{d}$, $e_{q}$ voltages and $i_{d}$, $i_{q}$ currents. The instantaneous real and reactive power in the d-q frame are given by \cite{main6}:

\begin{subequations}
	\begin{align}
	P = e_{d}i_{d} + e_{q}i_{q} \label{eq:Power_active}\\
	Q = e_{q}i_{d} - e_{d}i_{q} \label{eq:Power_reactive}
	\end{align}
\end{subequations}

From \cite{maineq}, the square of the output voltage $ V_{o}^2 $ in state space is given by

\begin{subequations}
	\begin{align}
	& x = V_{o}^2 & \label{eq:v22} \\
	& \dot{x} = -\frac{2}{R_{l}C}x + \frac{3e_{d}}{C}i_{d} + \frac{3e_{q}}{C}i_{q} & \label{eq:v2dott}
	\end{align}
\end{subequations}

Combining equations \ref{eq:Power_active}, \ref{eq:Power_reactive} and \ref{eq:v2dott}, the state space model of the active ($P$) and reactive power ($Q$) in the stationary d-q frame as given in \cite{main}. This considers that there is a power balance between the input and output side of the rectifier. The equation is given below:

\begin{gather}
\label{eq:matrixMain}
\begin{bmatrix} 
	\dot{V_{o}^2}\\[1em] 
	\dot{P} \\[1em] 
	\dot{Q} 
\end{bmatrix}
=
\begin{bmatrix} 
	-\frac{2}{R_{l}C} & \frac{3}{C} & 0 \\[1em]
	0 & -\frac{r_{s}}{L_{s}} & -\omega \\[1em]
	0 & \omega & \frac{r_{s}}{L_{s}} 
\end{bmatrix}
\begin{bmatrix} 
	V_{o}^2 \\[1em] P \\[1em] Q 
\end{bmatrix}
+
\begin{bmatrix} 
	0 & 0 \\[1em]
	\frac{1}{L_{s}} & -\frac{D_{d}}{L_{s}} \\[1em]
	0 & \frac{D_{q}}{L_{s}}
\end{bmatrix}
\begin{bmatrix} 
	e_{d} \\[1em]
	V_{o}
\end{bmatrix}
\end{gather} 

where $\omega = 2\pi\/f$ is the angular frequency, $f$ is the frequency of the three-phase system, $V_{o}$ is the DC output voltage. $D_{d}$ and $D_{q}$ are the d-q components of the duty cycle respectively. $L_{S} = \frac{L}{\sqrt{3}E}$ and $r_{S} = \frac{r_{L}}{\sqrt{3}E}$. The root mean square (RMS) value of the synchronous source voltages ($e_{a}$, $e_{b}$, $e_{c}$) is denoted by $E$.

In order to achieve decoupling of the active and reactive power which can be seen in equation \ref{eq:matrixMain}, \cite{main} decoupled it into two main parts as shown in equations \ref{eq:decouple1} and \ref{eq:decouple2}.

\begin{subequations}
	\begin{align}
	& \dot{P} = -\frac{r_{s}}{L_{s}}P + u_{p} & \label{eq:decouple1} \\ 
	& \dot{Q} = \frac{r_{s}}{L_{s}}Q  + u_{q} &
	\label{eq:decouple2}
	\end{align}
\end{subequations} 

where $ u_{p} = \dfrac{e_{d} - L_{s}\omega\/Q - D_{d}V_{o}}{L_{s}}$ and $ u_{q} = \dfrac{L_{s}\omega\\P + D_{q}V_{o}}{L_{s}}$ are designed as the virtual control signals for P and Q. This ensures that P and Q can be controlled independently with the help of $u_{p}$ and $u_{q}$ respectively. 

\section{Dynamic Model of the Output DC Voltage and Reactive Power}
Using equations \ref{eq:decouple1}, and \ref{eq:decouple2}, it can be seen that the active power $P$ and reactive power $Q$ can be controlled independently and by substituting \ref{eq:decouple1} into \ref{eq:matrixMain}, the output voltage $V_{o}$ can equally be controlled. Here, our goal is to control both the output voltage and the reactive power simultaneously. From these equations, the models of the output voltage and reactive power can be obtained. The derivations can be found in \cite{main}.

\subsection{Output Voltage Model}
By substituting equation \ref{eq:decouple1} into the derivative of $\dot{V_{o}^2}$ from equation \ref{eq:matrixMain}, the dynamic model of the DC output voltage is given as follows:

\begin{equation} 
\label{eq:voltdynamic}
\begin{split}
	\ddot{x}_{p}(t) & = a_{p}\dot{x}_{p}(t) + b_{p}u_{p}(t) + c_{p}z(t) \\
	 & = a_{pn}\dot{x}_{p}(t) + b_{pn}u_{p}(t) + c_{pn}z(t) + w_{1}(t) \\	
\end{split}
\end{equation}

where $x_{p}(t) = V_{o}^2$; $z(t) = P$; $a_{p} = \frac{-2}{R_{l}C}$;  $b_{p} = \frac{3}{C}$; and  $c_{p} = \frac{-3r_{s}}{L_{s}C}$.

\vspace{0.1in}
To account for the variations in the parameters, they are adjusted with their variations as follows;

\begin{equation} 
\label{eq:voltvariations}
\begin{split}
	a_{p} & = a_{pn} + \Delta\/a_{pn} \\
	b_{p} & = b_{pn} + \Delta\/b_{pn} \\
	c_{p} & = c_{pn} + \Delta\/c_{pn} \\	
\end{split}
\end{equation}   

where the nominal values of $a_{p}$, $b_{p}$ and $c_{p}$ are denoted by $a_{pn}$, $b_{pn}$, and $c_{pn}$ respectively. 

The system parameter variations are denoted by $\Delta\/a_{pn}$ , $\Delta\/b_{pn}$  and $\Delta\/c_{pn}$  respectively. These variations are lumped into the variable $w_{1}$ in equation \ref{eq:voltdynamic}. Therefore, $w_{1}$ is given in \ref{eq:w1}.

\begin{equation} 
\label{eq:w1}
w_{1}(t) = \Delta\/a_{pn}\dot{x}_{p}(t) + \Delta\/b_{pn}u_{p}(t) + \Delta\/c_{pn}z(t)
\end{equation}  

For ease of analysis, $w_{1}$ is assumed to be bounded by a positive constant $\rho\/_{p}$, i.e $\mid\/w_{1}(t)\mid < \rho\/_{p}$. 

\subsection{Reactive Power Model}
In similar manner, the dynamic model of the reactive power from equation \ref{eq:matrixMain} and \ref{eq:Power_reactive} can be deduced as follows \cite{main}:

\begin{equation} 
\label{eq:reacdynamic}
\begin{split}
\dot{x}_{q}(t) & = d_{q}x_{q}(t) + u_{q}(t) \\
& = d_{qn}x_{q}(t) + u_{q}(t) + g(t) \\	
\end{split}
\end{equation}

where $x_{q}(t) = Q$; and  $d_{q} = \frac{r_{s}}{L_{s}}$.

\vspace{0.1in}
Similarly, to account for the variations in the parameter $d_{q}$, it is adjusted with its variation as follows;

\begin{equation} 
\label{eq:reacvariation}
\begin{split}
d_{p} & = d_{qn} + \Delta\/d_{qn} 	
\end{split}
\end{equation}

where the nominal value of $d_{q}$ is denoted by $d_{qn}$. The parameter variation is denoted by $\Delta\/d_{qn}$. This variation is lumped into the variable $g(t)$ in equation \ref{eq:reacvariation}. Therefore, $g(t)$ is given in \ref{eq:g}.   

\begin{equation} 
\label{eq:g}
g(t) = \Delta\/d_{qn}\dot{x}_{q}(t)
\end{equation}  

Again, for ease of analysis, $g(t)$ is assumed to be bounded by a positive constant $\rho\/_{q}$, i.e $\mid\/g(t)\mid < \rho\/_{q}$. 

\section{Backstepping Controller (BSC) Design}
Backstepping control (BSC) for the output voltage $V_{o}$ and reactive power $Q$ model has been proposed in \cite{main}. Adaptive BSC for the output voltage stems from its BSC derivation.
 
\subsection{BSC Design for the Output voltage}
The design objective is for the square of the output voltage to follow a desired reference voltage under varying system uncertainties. The BSC is modeled as follows \cite{main};

Suppose the square of the output voltage is
\begin{equation} 
 V_{o}^2 = x_{p}
\end{equation}
 
and the square of the reference voltage is
\begin{equation} 
V_{o}^{*2} = x_{p}^*
\end{equation}

The tracking error $e_v$ and its derivative $\dot{e}_v$ are then given as:

\begin{subequations}
	\begin{align}
	& e_v = x_{p} - x_{p}^* & \label{eq:ev} \\ 
	& \dot{e}_v = \dot{x}_{p} - \dot{x}_{p}^* & \label{eq:evdot}
	\end{align}
\end{subequations} 

For BSC, we define a stabilizing function $\alpha$ and its derivative $\dot{\alpha}$ as:

\begin{subequations}
	\begin{align}
	& \alpha = \dot{x}_{p}^* - k_{v}e_{v} & \label{eq:alpha} \\ 
	& \dot{\alpha} = \ddot{x}_{p}^* - k_{v}\dot{e}_{v} & \label{eq:alphadot}
	\end{align}
\end{subequations} 

where $k_v$ is a positive constant. The derivative of the square of the output voltage $\dot{x}_{p}$ can also be termed the virtual control variable. Therefore, we can define the virtual control error $e_s$ and its derivative $\dot{e}_s$ as follows:

\begin{subequations}
	\begin{align}
	& e_s = \dot{x}_{p} - \alpha & \label{eq:es} \\ 
	& \dot{e}_s = \ddot{x}_{p} - \dot{\alpha} & \label{eq:esdot}
	\end{align}
\end{subequations} 

Since it is desired that the error $e_v$ tend to zero, the first potential Lyapunov function is given as:

\begin{equation} 
\label{eq:v1}
V_{1} = \frac{1}{2}e_{v}^2
\end{equation} 

The derivative of $V_{1}$ is then

\begin{equation} 
\label{eq:v1dot}
\begin{split}
\dot{V}_{1}  = e_{v}\dot{e}_{v} & = e_{v}(\dot{x}_{p} - \dot{x}_{p}^*)  \\
& = e_{v}(\dot{x}_{p} - \alpha - k_{v}e_{v})  \\
& = e_{v}e_{s} - k_{v}e^2_{v}	
\end{split}
\end{equation}

Substituting \ref{eq:voltdynamic} into \ref{eq:esdot} gives

\begin{equation} 
\label{eq:esdot2}
\dot{e}_s = a_{pn}\dot{x}_{p} + b_{pn}u_{p} + c_{pn}z + w_{1} - \dot{\alpha}
\end{equation} 

Looking at \ref{eq:esdot2}, if we take the control signal $u_p$ to be of the following form which would make \ref{eq:esdot2} to be in terms of only the error variables,

\begin{equation} 
\label{eq:up}
u_p = \dfrac{1}{b_{pn}}\left\{ -a_{pn}\dot{x}_{p} - c_{pn}z - w_{1} + \dot{\alpha} -k_{s}e_{s} - e_{v} \right\} 
\end{equation}

where $k_s$ is a positive constant. $\dot{e}_s$ becomes:

\begin{equation} 
\label{eq:esdot3}
\begin{split}
\dot{e}_s = -k_{s}e_{s} - e_{v}
\end{split}
\end{equation}

To enforce $e_s$ to zero, the second potential Lyapunov function follows from \ref{eq:esdot3} as:

\begin{equation} 
\label{eq:v2}
V_{2} = \frac{1}{2}e_{v}^2 + \frac{1}{2}e_{s}^2
\end{equation} 

The derivative of $V_{2}$ is then

\begin{equation} 
\label{eq:v2dot}
\begin{split}
\dot{V}_{2} & = \dot{V}_{1} + e_{s}\dot{e}_{s} \\
& = e_{v}e_{s} - k_{v}e^2_{v} + e_{s}(-k_{s}e_{s} - e_{v})	\\
& = - k_{v}e^2_{v} - k_{s}e^2_{s}
\end{split}
\end{equation}

From \ref{eq:v2dot}, it can be seen that $\dot{V}_{2}$ is negative definite. This implies that the errors $e_v$ and $e_s$ will go asymptotically to zero. This ensures that with the control signal $u_p$ above, the output voltage's stability is guaranteed.

\subsection{BSC Design for the Reactive Power}
The design objective here is same as for the output voltage. We want the reactive power $Q$ to follow a desired reference reactive power under varying system uncertainties. Again, the BSC is modeled as follows \cite{main};

Suppose the square of the reactive power is denoted by
\begin{equation} 
Q = x_{q}
\end{equation}

and the square of the reference voltage 
\begin{equation} 
Q^{*} = x_{q}^*
\end{equation}

The tracking error $e_q$ and its derivative $\dot{e}_q$ are then given as:

\begin{subequations}
	\begin{align}
	& e_q = x_{q} - x_{q}^* & \label{eq:eq} \\ 
	& \dot{e}_q = \dot{x}_{q} - \dot{x}_{q}^* & \label{eq:eqdot}
	\end{align}
\end{subequations} 

We want $e_q$ to go to zero, hence, the third potential Lyapunov function is:

\begin{equation} 
\label{eq:v3}
V_{3} = \frac{1}{2}e_{q}^2
\end{equation}  

The derivative of $V_{3}$ is as below after substituting \ref{eq:reacdynamic}:

\begin{equation} 
\label{eq:v3dot}
\begin{split}
\dot{V}_{3}  = e_{q}\dot{e}_{q} & = e_{q}(\dot{x}_{q} - \dot{x}_{q}^*)  \\
& = e_{q}(d_{qn}x_{q} + u_{q} + g - \dot{x}_{q}^*)  
\end{split}
\end{equation}

In order to make $\dot{V}_{3}$ negative definite, i.e $\dot{V}_{3} = - k_{q}e_{q}^2$, where $k_{q}$ is a positive constant, the control signal $u_q$ can be selected as follows:

\begin{equation} 
\label{eq:uq}
u_q = -d_{qn}x_{q} + u_{q} - k_{q}e_{q} + \dot{x}_{q}^* - g
\end{equation}

This selection of $u_q$ will ensure that the reactive power error $e_q$ goes asymptotically to zero. This ensures the stability of the reactive power as well.

\section{Adaptive-BSC Design for the Output voltage}
The previous section showed in detail how the BSC is used to achieve stability and ensure that the output voltage and reactive power follow the desired voltage and power. These follow, keenly, the derivations from \cite{main}. However, the assumption above is that the values of the components are ideal and known, giving rise to little variation.

However, in practice, not all nominal values are available to the controller before hand to perform analysis. Our system must be robust enough to cater for unknown values and still behaves as expected. To be able to achieve this feat, an adaptive-BSC approach is proposed. Here, we assume that the exact value of the output load resistance value $R_l$ is unknown to the controller (meaning that $R_l$ can have any value). This will affect the way in which the controller would react to the system. Since $R_l$ can vary, it follows that $a_{pn} = \frac{-2}{R_{l}C}$ is also varying and its exact value cannot be ascertained.

Hence, the system can only perform an estimation of $a_{pn}$ based on some adaptation law. This law would seek to minimize the difference between the actual unknown value of $a_{pn}$ and the system estimated value. The control variable $u_p$ would be affected by the fact that $a_{pn}$ is varying since its value depend on it as seen in equation \ref{eq:up}. 

The new adaptive control signal $u_p$ and the adaptation law is derived as follows:

\vspace{0.2 in}

Let $\mathbf{\hat{a}_{pn}}$ be the system estimated value of $a_{pn}$ and $\mathbf{\tilde{a}_{pn}}$ be the difference between $a_{pn}$ and $\hat{a}_{pn}$.

\begin{equation} 
\label{eq:atail}
\tilde{a}_{pn} = a_{pn} - \hat{a}_{pn}
\end{equation}

$u_p$ from equation \ref{eq:up} is then modified adaptively as:

\begin{equation} 
\label{eq:upadap}
u_p = \dfrac{1}{b_{pn}}\left\{ -\hat{a}_{pn}\dot{x}_{p} - c_{pn}z - w_{1} + \dot{\alpha} -k_{s}e_{s} - e_{v} \right\} 
\end{equation}

where all other variables retain their earlier meanings. $\dot{e}_s$ then gives

\begin{equation} 
\label{eq:esdot4}
\begin{split}
\dot{e}_s & = \ddot{x}_{p} - \dot{\alpha} \\
& = (a_{pn} - \hat{a}_{pn})\dot{x}_{p} - k_{s}e_{s} - e_{v} \\
& = \tilde{a}_{pn}\dot{x}_{p} - k_{s}e_{s} - e_{v}
\end{split}
\end{equation}

Again, to enforce $e_s$ to zero, the potential Lyapunov function $V_4$ is given as follows:

\begin{equation} 
\label{eq:v4}
V_{2} = \frac{1}{2}e_{v}^2 + \frac{1}{2}e_{s}^2 + \frac{1}{2\gamma}\tilde{a}_{pn}^2
\end{equation} 

The derivative of $V_{4}$ is then

\begin{equation} 
\label{eq:v4dot}
\begin{split}
\dot{V}_{4} & = \dot{V}_{1} + e_{s}\dot{e}_{s} + \frac{1}{\gamma}\tilde{a}_{pn}\dot{\tilde{a}}_{pn}  \\
& = e_{v}e_{s} - k_{v}e^2_{v} + e_{s}(\tilde{a}_{pn}\dot{x}_{p} - k_{s}e_{s} - e_{v}) + \frac{1}{\gamma}\tilde{a}_{pn}\dot{\tilde{a}}_{pn} \\
& = - k_{v}e_{v}^2 - k_{s}e_{s}^2 + \tilde{a}_{pn}\big(\dot{x}_{p}e_{s} + \frac{1}{\gamma}\dot{\tilde{a}}_{pn}\big)
\end{split}
\end{equation}

where $\gamma$ is the adaptation gain.

To ensure that $V_{4}$ is negative definite which would ensure that $e_s$ goes to zero asymptotically, the last term of equation \ref{eq:v4dot} must be equal to zero. This means that:

\begin{equation}
\label{eq:adaprule}
\dot{\tilde{a}}_{pn} = -\gamma\/\dot{x}_{p}e_{s}
\end{equation}

Equation \ref{eq:adaprule} is the adaptation rule which would ensure that the system estimate $\hat{a}_{pn}$ is as close as possible to $a_{pn}$. Control signal $u_p$ obtained in \ref{eq:upadap} is the modified adaptive control signal which would ensure $V_o$ follows the desired $V_{o}^*$.

\section{Results and Simulations}
The performance of both the BSC proposed in \cite{main} and the adaptive-BSC proposed here were verified using MATLAB. MATLAB codes for the dynamic equations of the system developed from equation \ref{eq:matrixMain} to \ref{eq:adaprule} were developed. Table \ref{tab:parameters} shows the parameters used for the simulation. For the parts requiring differentials, Euler's approximation of a differential equation was used with a step size ($h = 10^{-4}$). 

\begin{table}[ht!]
	\centering
	\label{tab:parameters}%
	\caption{Simulation Parameters}
	\begin{tabular}{ccc}
		\toprule
		\textbf{Parameter} & \textbf{Symbol} & \textbf{Value}  \\
		\midrule
		\vspace{0.05in}
		Utility Phase Voltage (RMS)  & E & 311V \\
		\vspace{0.05in}
		Utility Frequency & F & 60Hz \\
		\vspace{0.05in}
		Nominal Inductance  & Ln & 12mH \\
		\vspace{0.05in}
		Source Resistance & $r_L$ & $0.1\Omega$ \\ 
		\vspace{0.05in}
		Filtering Capacitance & $C$ & 3.3mF \\ 
		\vspace{0.05in}
		Nominal Output Load & $R_l$ & $200\Omega$ \\ 
		\vspace{0.05in}
		Active Power & P & 5kW \\
		\vspace{0.05in}
		Positive Constants & $k_v$, $k_s$, and $k_q$ & 500, 500, and 0.2 \\ 
		\vspace{0.05in}
		Positive Constants & $\rho_p$ and $\rho_q$ & 0.5 and 0.5 \\ 
		\vspace{0.05in}
		Adaptive gain & $\gamma$ & $10^{-3}$ \\ 
		\bottomrule
	\end{tabular}%
\end{table}%

For both simulations (BSC only and Adaptive-BSC), a simulation time of \textbf{2.5} seconds was used with a step size same as $h$. A reference voltage ($V_{o}^*$) of 1000V is set at the beginning and then changes to 800V at 50\% simulation time. A reference reactive power ($Q^*$) of 0kVAr was set at the beginning of the simulation and changed to 5kVAr at about 70\% simulation time.

\subsection{Results With BSC Only}
\begin{figure}[htbp!]
	\centering
	\includegraphics[width=0.45\textwidth]{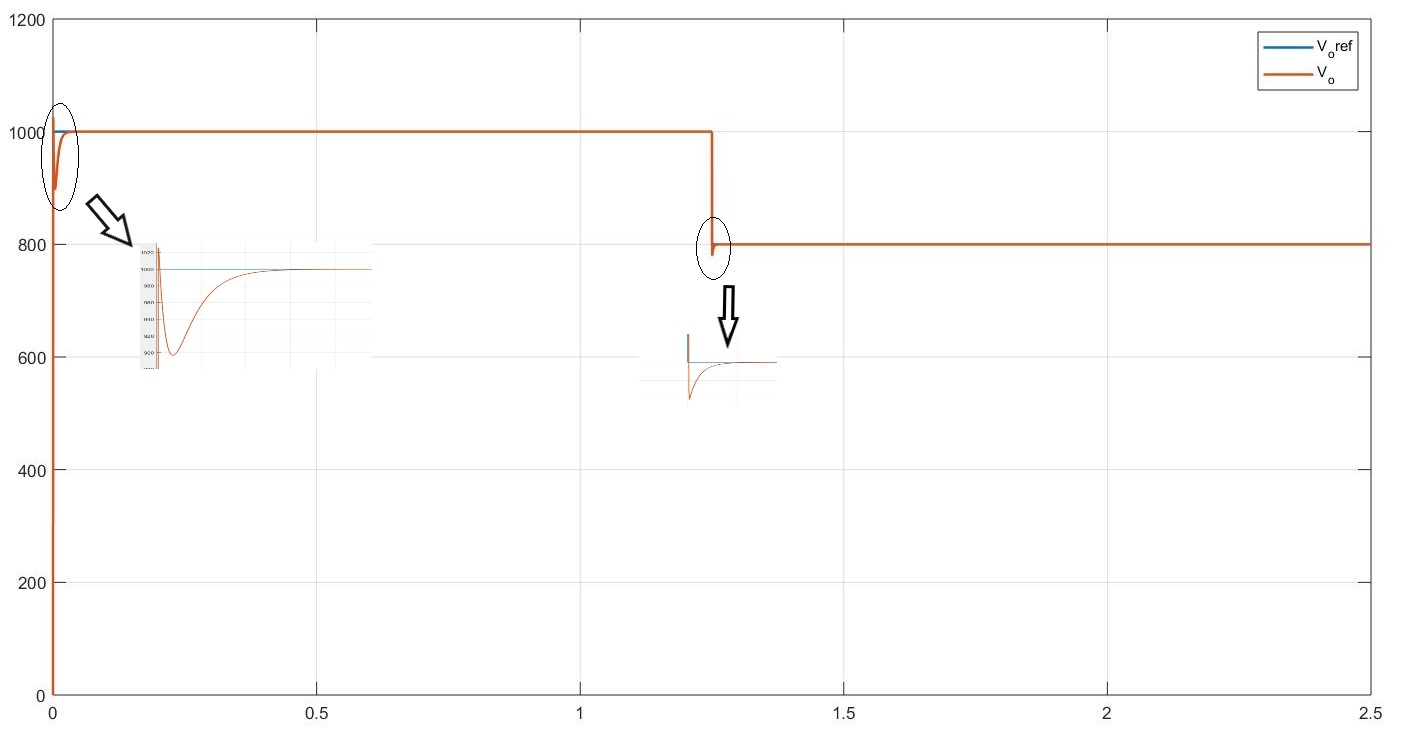}
	\caption[]{Output Voltage $V_o$ and $V_{o}^*$ via BSC}
	\label{fig:Vbsc}
\end{figure}

Figure \ref{fig:Vbsc} shows both the simulated system output voltage $V_o$ (in Orange) and the reference voltage $V_{o}^*$ (in Blue). Throughout the simulation, $V_{o}^*$ simply run as described above. At the start of the simulation, however, a little jump of about 10V higher than the reference of 1000V can be noticed in the system output. This is due to the fact that the system is just starting out and the BSC controller is starting from 0V. The system then settles out in about 0.05 sec. This is evident in the zoomed in portion shown on the figure at the start. 

At about 50\% of the simulation time, a sharp decline can be noticed in the reference voltage from 1000V to 800V. Again, the controller can be seen to follow this trajectory and it achieved this in less than 0.002 sec. This can be seen in the zoomed in part of figure \ref{fig:Vbsc}.

\begin{figure}[htbp!]
	\centering
	\includegraphics[width=0.45\textwidth]{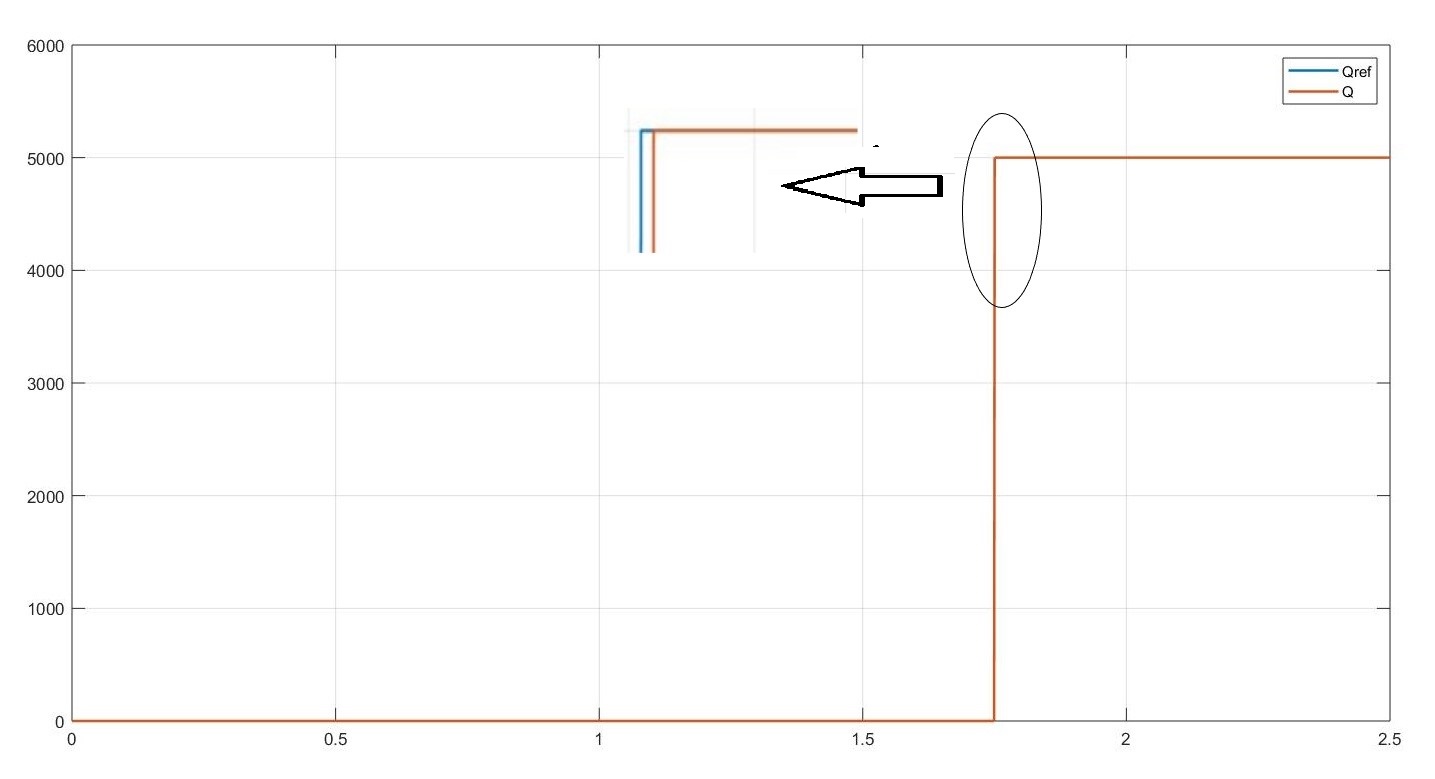}
	\caption[]{Reactive Power $Q$ and $Q^*$ via BSC}
	\label{fig:Qbsc}
\end{figure}

Also, the system reactive power $Q$ can be seen in figure \ref{fig:Qbsc} to follow the desired reactive power $Q^*$. This tracking can be observed to be very rapid with almost 0 percentage overshoot. This can be attributed to the fact that the control signal for $Q$ ($u_q$) has far fewer variables it depends on.  

\subsection{Results With Adaptive-BSC}
\begin{figure}[htbp!]
	\centering
	\includegraphics[width=0.45\textwidth]{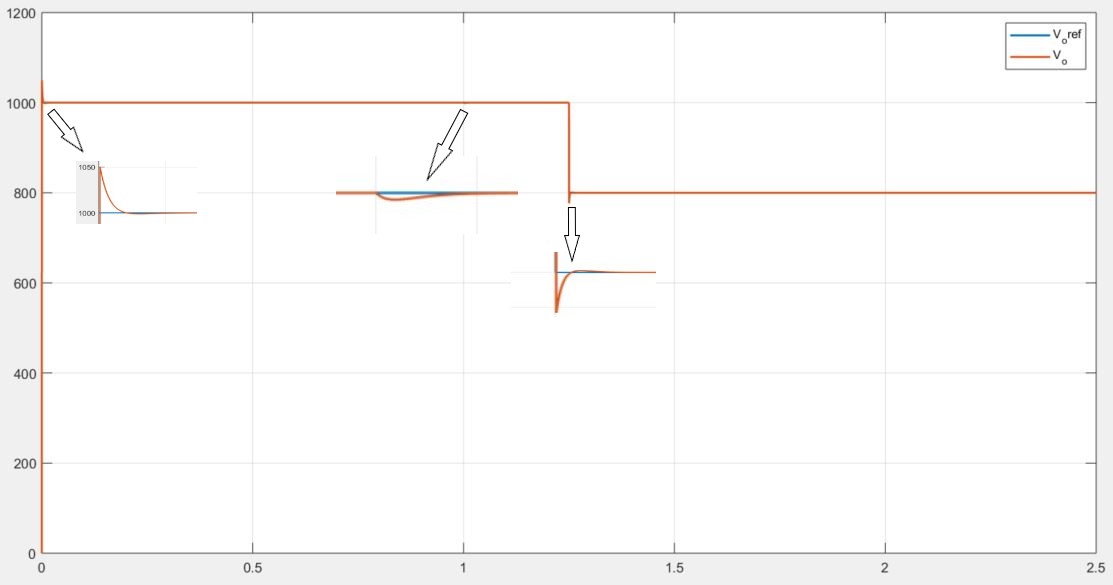}
	\caption[]{Output Voltage $V_o$ and $V_{o}^*$ via Adaptive-BSC}
	\label{fig:Vadap}
\end{figure}

Adaptive-BSC simulation of the output voltage $V_o$ and the desired output $V_{o}^*$ is as shown in \ref{fig:Vadap}. Here, the system has been subjected to tougher variations compared to the basic BSC described above. This is so because the actual nominal value of the load $R_l$ remains unknown to the controller because it was varied unlike for the BSC where everything is known. The controller therefore, needed to perform more work under a tougher uncertainty conditions.

\begin{figure}[htbp!]
	\centering
	\includegraphics[width=0.45\textwidth]{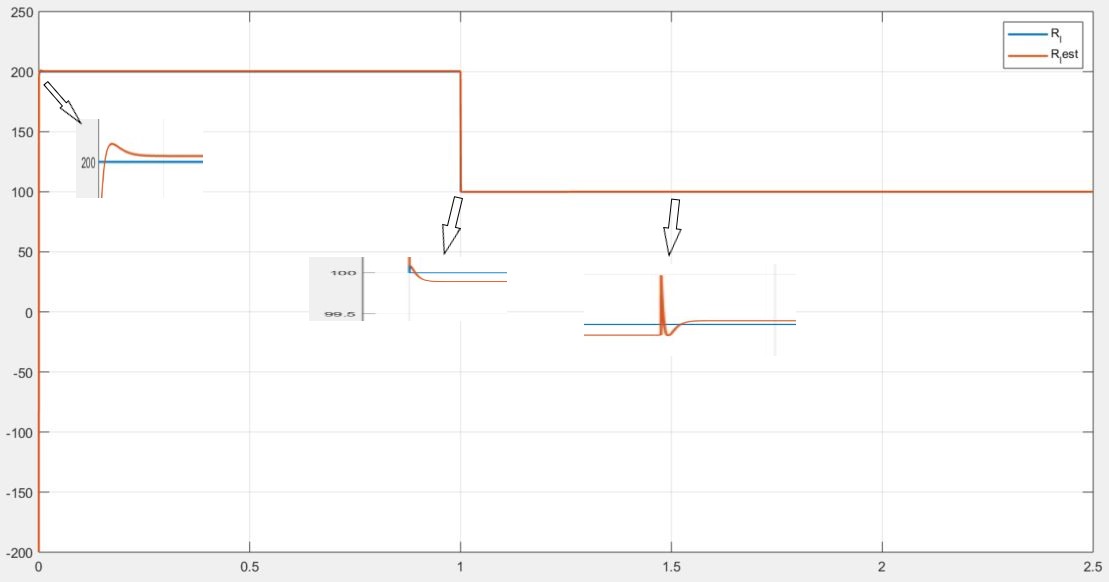}
	\caption[]{System Estimated Load $R_{est}$ and True Load $R_l$ via Adaptive-BSC}
	\label{fig:Radap}
\end{figure}

The load was varied from 200 $\Omega$ at the start to 100 $\Omega$ at about 1 sec simulation time as shown in figure \ref{fig:Radap}. The controller, therefore, needed to estimate the value of $R_l$ as it is unknown to it. The estimation curve is shown along with the tested nominal value in \ref{fig:Radap}. It should be noted that any changing value of $R_l$ would produce a similar result.

Looking again at the output voltage in figure \ref{fig:Vadap}, we can see how the controller tries to adjust itself to follow the reference when system parameter changes (at the start of simulation, at about 1 sec when $R_l$ changed and at about 50\% simulation time when the reference voltage changed).  

Comparing the adaptive-BSC with what was obtained with the BSC, it can be seen that the adaptive-BSC performed better than the BSC in all forms of evaluation. From the zoomed version of \ref{fig:Vadap}, at the start of the simulation, it took less time ()about 0.005 sec) and with no undershoot for the controller to produce perfect reference following as compared with 0.05 sec of the ordinary BSC. This phenomenon can also be observed when the reference voltage changed from 1000V to 800V as it took the adaptive controller about 0.005 sec with no overshoot. Although, the BSC also produced no overshoot, but it took longer time to achieve this (about 0.05 sec).  

It should also be noted that the adaptive-BSC worked under more challenging system variations as compared with the BSC, yet, it performed 10 times better in terms of settling time than the BSC.

\subsection{Adjustment to the Main Equations}
To come up with the adaptive smulation, little adjustments were made to the adaptive controller code. The main adjustment was in the calculation of the adaptive rule which we got analytically as $\dot{\tilde{a}}_{pn} = -\gamma\/\dot{x}_{p}e_{s}$ but, in code, $\dot{\tilde{a}}_{pn} = -\gamma\/e_{v}$ was used. Also, for the calculation of $u_p$, $\tilde{a}_{pn}$ was used instead of $\hat{a}_{pn}$.

\section{Conclusions}
In this study, we have successfully designed an adaptive backstepping control (Adaptive-BSC) system for a three-phase PWM rectifier. MATLAB simulation was performed to verify the proposed controller. The results were compared to its BSC counterpart. Using the developed dynamic equations, MATLAB code was written and simulated for a period of 2.5 sec. It was desired for the output voltage to track a reference voltage and the output reactive power to track a reference reactive power. Both the BSC and adaptive-BSC were investigated. The adaptive controller had to estimate the value of the load adaptively unlike the BSC system which is aware of the constant load from onset. The proposed adaptive-BSC worked under more challenging system variations as compared with the BSC, yet, it performed 10 times better in terms of settling time than the BSC. The designed system can be used for arbitrary DC voltage regulation which might be needed for the control of industrial systems.

\balance


\begin{thebibliography}{1}
\bibitem{kant}
K. Kant and C. Jain, \emph{"A hybrid diesel-wind-PV-based energy generation system with brushless generators"}, IEEE Trans. Ind. Informatics, vol. 13, no. 4, pp. 1714-1722, 2017.

\bibitem{main}
R. Wai, \emph{"Design of Backstepping Control for Direct Power Control of Three-Phase PWM Rectifier,"} 2018 3rd Int. Conf. Intell. Green Build. Smart Grid, pp. 1-4.

\bibitem{maineq}
W. Zhang, Y. Hou, X. Liu, and Y. Zhou, \emph{"Switched Control of Three-Phase Voltage Source PWM Rectifier Under a Wide-Range Rapidly,"} IEEE Trans. Power Electron., vol. 27, no. 2, pp. 881–890, 2012.

\bibitem{dpc1}
J. Ge, Z. Zhao, L. Yuan, T. Lu, and F. He, \emph{"Direct Power Control Based on Natural Switching Surface for Three-Phase,"} IEEE Trans. Power Electron., vol. 30, no. 6, pp. 2918–2922, 2015.

\bibitem{dpc2}
W. P. Loop, J. Ma, S. Member, W. Song, and S. Jiao, \emph{"Power Calculation for Direct Power Control of Single-Phase Three-Level Rectifiers,"} IEEE Trans. Ind. Electron., vol. 63, no. 5, pp. 2871–2882, 2016.

\bibitem{dpc3}
Z. Song, Y. Tian, S. Member, and Z. Yan,\emph{"Direct Power Control for Three-Phase Two-Level Voltage-Source Rectifiers Based on Extended-State Observation,"} IEEE Trans. Ind. Electron., vol. 63, no. 7, pp. 4593–4603, 2016.

\bibitem{dpc4}
J. Wang, \emph{"Modified SVPWM-Controlled Three-Port Three-Phase AC - DC Converters With Reduced Power Conversion Stages for Wide Voltage Range Applications,"} IEEE Trans. Power Electron., vol. 33, no. 8, pp. 6672–6686, 2018.

\bibitem{main2}
S. Weerasinghe, U. K. Madawala, and D. J. Thrimawithana, \emph{"A matrix converter-based bidirectional contactless grid interface"}, IEEE Trans. Power Electron., vol. 32, no. 3, pp. 1755-1766, 2017.

\bibitem{main3}
M. Yilmaz and P.T. Krein, \emph{"Review of the impact of vehicle-to-grid technologies on distribution systems and utility interfaces"}, IEEE Trans. Power Electron., vol. 28, pp. 5673-5689, 2013.

\bibitem{main4}
X. Zhang and C. W. Zhang, \emph{"PWM rectifier and control"}, Bei Jing: China Machine Press, 2011.

\bibitem{main5}
Y. Cho and K. B. Lee, \emph{"Virtual-flux-based predictive direct power control of three-phase PWM rectifiers with fast dynamic response"}, IEEE Trans. Power Electron., vol 31, no. 4, pp. 3348-3359, 2016.

\bibitem{main6}
H. Chen, A. Prasai, and D. Divan, \emph{"A modular isolated topology for instantaneous reactive power compensation"}, IEEE Trans. Power Electron., vol. 33, no. 2, pp. 975-986, 2018.

\bibitem{main8}
N. M. Dehkordi, N. Sadati, and M. Hamzeh, \emph{"A robust backstepping highorder sliding mode control strategy for grid-connected DG units with harmonic/interharmonic current compensation capability"}, IEEE Trans. Sustainable Energy, vol. 8, no. 2, pp. 561-572, 2017.

\bibitem{main7}
R. J. Wai, J. X. Yao, and J. D. Lee, \emph{"Backstepping fuzzy-neuralnetwork control design for hybrid maglev transportation System"}, IEEE Trans. Neural Networks and Learning Systems, vol. 26, no. 2, pp. 302 - 317, 2015.

\bibitem{adap1}
T. K. Roy, S. Member, A. Mahmud, S. Member, and A. M. T. Oo, \emph{"Robust Adaptive Backstepping Excitation Controller Design for Higher-Order Models of Synchronous Generators in Multimachine Power Systems,"} IEEE Trans. Power Systems, vol. 34, no. 1, pp. 40–51, 2019.

\bibitem{adap2}
J. Cai, C. Wen, H. Su, Z. Liu, and L. Xing, \emph{"Adaptive Backstepping Control for a Class of Nonlinear Systems With Non-Triangular Structural Uncertainties,"} IEEE Trans. Auto. Ctrl., vol. 62, no. 10, pp. 5220–5226, 2017.

\bibitem{adap3}
L. Sun, S. Tong, and Y. Liu, \emph{"Adaptive Backstepping Sliding Mode Control of Static Var Compensator,"} IEEE Trans. Ctrl. Sys. Tech., vol. 19, no. 5, pp. 1178–1185, 2011.

\bibitem{adap4}
G. Wang, R. Wai, and Y. Liao, \emph{"Design of backstepping power control for grid-side converter of voltage source converter-based high-voltage dc wind power generation system,"} IET Renew. Power Gener., vol. 7, no. December 2012, pp. 118–133, 2013.

\bibitem{adap5}
T. K. Roy, S. Member, and A. Mahmud, \emph{"Dynamic Stability Analysis of Hybrid Islanded DC Microgrids Using a Nonlinear Backstepping Approach,"} IEEE Syst. J., vol. 12, no. 4, pp. 3120–3130, 2018.

\bibitem{adap6}
T. K. Roy et al., \emph{"Nonlinear Adaptive Backstepping Controller Design for Islanded DC Microgrids,"} IEEE Trans. Ind. Appl., vol. 54, no. 3, pp. 2857–2873, 2018.

\end{thebibliography}
\end{document}